\documentclass[%
 aip,
 amsmath,amssymb,
 reprint,%
]{revtex4-1}

\usepackage{graphicx}
\usepackage{dcolumn}
\usepackage{bm}

\usepackage[utf8]{inputenc}
\usepackage[T1]{fontenc}
\usepackage{mathptmx}
\usepackage{hyperref}
\usepackage{textcomp}
\usepackage{makecell}
\usepackage[english]{babel}
\usepackage{blindtext}
\usepackage{color}
\usepackage{soul}
\usepackage{placeins}
\usepackage{siunitx}
\usepackage{changes}

\definechangesauthor[name={Per cusse}, color=orange]{per}

\begin{document}

\preprint{AIP/123-QED}

\title[Thermal activation of valley-orbit states of neutral magnesium in silicon]{Thermal activation of valley-orbit states of neutral magnesium in silicon}

\author{R. J. S. Abraham}
\affiliation{Department of Physics, Simon Fraser University, Burnaby, British Columbia, Canada V5A 1S6}
\author{V. B. Shuman}
\affiliation{Ioffe Institute, Russian Academy of Sciences, 194021 St. Petersburg, Russia}
\author{L. M. Portsel}
\affiliation{Ioffe Institute, Russian Academy of Sciences, 194021 St. Petersburg, Russia}
\author{A. N. Lodygin}
\affiliation{Ioffe Institute, Russian Academy of Sciences, 194021 St. Petersburg, Russia}
\author{Yu. A. Astrov}
\affiliation{Ioffe Institute, Russian Academy of Sciences, 194021 St. Petersburg, Russia}
\author{N. V. Abrosimov}
\affiliation{Leibniz-Institut f$\ddot{\text{u}}$r Kristallz$\ddot{\text{u}}$chtung (IKZ), 12489 Berlin, Germany}
\author{S. G. Pavlov} 
\affiliation{Institute of Optical Sensor Systems, German Aerospace Center (DLR), 12489 Berlin, Germany}
\author{H.-W. H$\ddot{\text{u}}$bers}
\affiliation{Institute of Optical Sensor Systems, German Aerospace Center (DLR), 12489 Berlin, Germany}
\affiliation{Humboldt Universit$\ddot{\text{a}}$t zu Berlin, Department of Physics, 12489 Berlin, Germany}
\author{S. Simmons}
\affiliation{Department of Physics, Simon Fraser University, Burnaby, British Columbia, Canada V5A 1S6}
\author{M. L. W. Thewalt}
\email[Corresponding author: ]{thewalt@sfu.ca}
\affiliation{Department of Physics, Simon Fraser University, Burnaby, British Columbia, Canada V5A 1S6}

\date{\today}

\begin{abstract}
Interstitial magnesium acts as a moderately deep double donor in silicon, and is relatively easily introduced by diffusion. Unlike the case of the chalcogen double donors, the binding energies of the even-parity valley-orbit excited states 1sT$_2$ and 1sE have remained elusive. Here we report on temperature dependence absorption measurements focusing on the neutral charge species. Our results demonstrate thermal activation from the ground state 1sA to the valley-orbit states, as observed by transitions from the thermally populated levels to the odd-parity states 2p$_0$ and 2p$_{\pm}$.
\end{abstract}

\maketitle

%

Interstitial magnesium (Mg$_i$) in silicon is a deep double donor defect that has been the subject of several past investigations \cite{Franks1967,Ho1972,Ho1993,Ho1998,Ho2003,2Ho2003,Ho2006} which have uncovered many features of the neutral (Mg$_i^0$) and singly ionized (Mg$_i^+$) species. Recent investigations \cite{Abraham2018, 2Abraham2018} have provided more detail and revealed new complexity, including information about complexes Mg forms with other defects. Mg incorporates into Si primarily as an interstitial defect, and inhabits  the tetrahedral (T$_d$) interstitial site \cite{Franks1967, Ho1972}. Relative to many single and double donor impurities in Si, such as the Group-V shallow single donors and Group-VI deep donors, the moderately deep double donor Mg$_i$ center has not been studied as comprehensively. While transitions to its odd parity excited states have been studied in some detail, there have been relatively few investigations of the even parity excited states. A thorough understanding of the electronic structure of Mg$_i$ may prove important in the design of semiconductor lasers using intracenter electronic transitions \cite{Pavlov2018, Hubers2005, Shastin2019}. Transitions to even parity excited states in particular are important to a complete understanding of cascade relaxation processes \cite{Ascarelli1961}. In such processes, electrons are initially captured by highly excited states and slowly diffuse to the impurity ground state through the ladder of available energy levels. Phonon(s) are emitted at each step of the ladder as described by Lax et al. \cite{Lax1960}. \par

A proposal by Morse et al. \cite{Morse2017} has recently suggested the use of deep donor centres in silicon (Si) as the basis for a new spin qubit-photonic cavity technology. The proposed technology relies on transitions from the ground state to the valley orbit level 1sT$_2$ which are forbidden within the effective mass theory (EMT) approximation. These transitions can be very strong for exceptionally deep donors such as the chalcogens \cite{Morse2017,Steger2009}, however no sign of these levels has been seen in absorption for Mg in Si until very recently \cite{Pavlov2018}. Mg is a double donor intermediate in depth between the shallow donors of Group-V and the deep-double donors of Group-VI. It may therefore lack the exceptionally strong central cell potential that allows for EMT-forbidden optical transitions between 1sA and 1sT$_2$. Electronic dipole transitions between 1sA and 1sE are in addition forbidden by parity, though potentially visible in Raman scattering. However, previous experiments have not revealed the presence of such transitions in Raman spectra of Si:Mg samples \cite{Abraham2018}. \par

Electronic transitions involving valley orbit excited states have been studied in past investigations of shallow donors like P, As, Sb, and Bi \cite{Mayur1993}. By thermally populating the 1sE and 1sT$_2$ states the authors were able to observe transitions to higher hydrogenic states such as 2p$_0$ and 2p$_{\pm}$. This technique is of course not viable for studies of very deep donors, and Pavlov et al. \cite{Pavlov2018} suggest that, as the shallowest of the deep donors, Mg$_i^0$ may be close to the limit of its viability. \par

Preliminary results discussed by Pavlov et al. \cite{Pavlov2018} have suggested binding energies of $\SI{47.4 \pm 0.1}{meV}$ and $\SI{49.9 \pm 0.2}{meV}$ for the 1sE and 1sT$_2$ levels of Mg$_i^0$, respectively, by observing weak transitions between these levels and 2p$_{\pm}$. An additional estimate by Pavlov et al. \cite{Pavlov2018} was calculated from the absorption spectra of Si:Mg samples subjected to a uniaxial strain. These measurements suggested a binding energy for 1sT$_2$ of $\SI{49.8 \pm 0.1}{meV}$ in the unstrained sample. Previous work by Ho and Ramdas \cite{Ho1972} established an empirical estimate of the the 1sA to 1sE transition energy for Si:Mg of $\SI{56.24}{meV}$ based on piezospectroscopic measurements. Binding energies determined in this work and by Pavlov et al. \cite{Pavlov2018} are somewhat lower than this value. Exceptionally deep donors, e.g. S, Se, Te, are known to have even smaller 1sE binding energies of $\sim\SI{31}{meV}$ meaning their 1s valley-orbit states are more hydrogen-like than helium-like. As the shallowest known isolated double donor, Mg is thought to be more helium-like than the deeper double donor impurities. As such, its somewhat deeper binding energies of the 1sT$_2$ and 1sE valley orbit excited states relative to S, Se, and Te, closer to the estimate of Ho and Ramdas \cite{Ho1972}, is not unexpected. In this study we elaborate on the work of Pavlov et al. \cite{Pavlov2018}, demonstrating that the valley-orbit states 1sT$_2$ and 1sE of Mg$_i^0$ may can indeed be thermally populated, leading to clearly resolved transitions from both valley-orbit states to the odd parity states 2p$_{0}$ and 2p$_{\pm}$. \par

In this study we measure the absorption spectra of a float-zone grown Si sample diffused with Mg. Parameters for Si:Mg sample preparation have been discussed in detail by Shuman et al. \cite{1Shuman2017,2Shuman2017}. Here we work with a lightly compensated Si sample with a boron concentration of approximately 1$\times$10$^{13}$ cm$^{-3}$, corresponding to the $^\textrm{nat}$Si low boron (LB) sample studied in our previous work \cite{Abraham2018}. \par

This sample was not selected for maximum Mg$_i^0$ concentration, as was done in the preliminary \cite{Pavlov2018} study but rather for the highest ratio of the desired Mg$_i^0$ absorption as compared to other absorption transitions in the 30 to $\SI{45}{meV}$ region. Many weak transitions in this region are as-yet unidentified, and can obscure the relatively weak transitions of interest. \par

All absorption measurements were performed using a Bruker IFS 125HR Fourier transform infrared (FTIR) spectrometer. A coated Mylar beam-splitter was used, along with a 4.2 K silicon bolometer with an 800 cm$^{-1}$ low-pass cold filter. Samples were mounted on a temperature-regulated stage inside a liquid helium cryostat with polypropylene windows. \par

\begin{figure}[htp]
\includegraphics[width=0.42\textwidth]{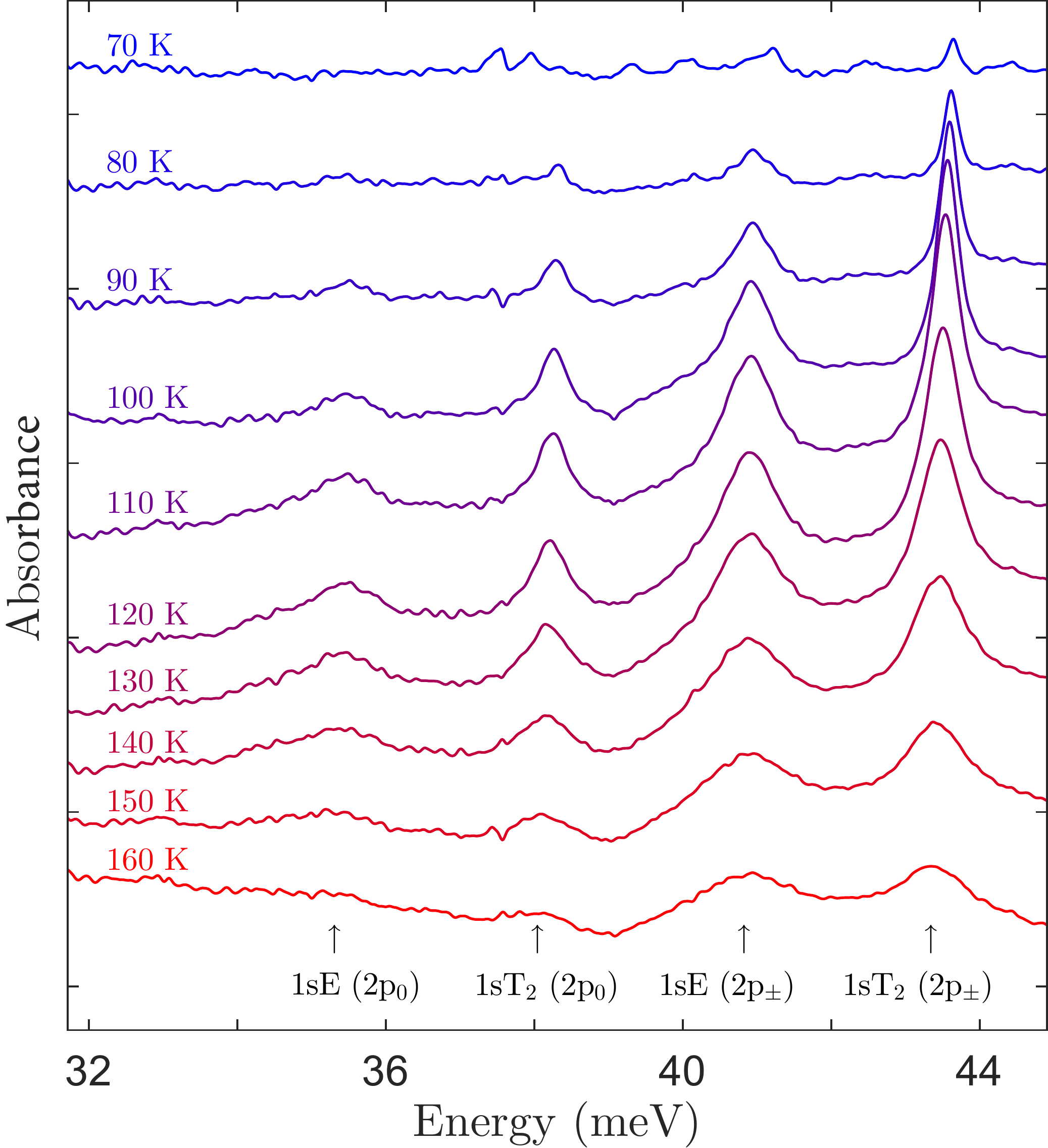}
\caption{Here we show absorbance spectra of Mg$_i^0$ over a range of temperatures. As temperature increases we see transitions from the thermally populated 1sE and 1sT$_2$ levels to 2p$_{\pm}$, corresponding to the higher energy pair of peaks and to 2p$_0$, corresponding to the lower energy pair. Spectra were collected at a resolution of $\SI{0.5}{cm^{-1}}$ ($\sim\SI{0.062}{meV}$).}
\label{1sT2_1sE_ThermalActivation}
\end{figure}

In Fig.~\ref{1sT2_1sE_ThermalActivation} we see thermally-induced absorbance resulting from transitions between the valley-orbit excited states and the higher-lying odd parity states 2p$_{0}$ and 2p$_{\pm}$. Some processing was necessary to remove an Si phonon feature corresponding to TO$_X$-TA$_X$. This was noted experimentally at $\SI{39.04}{meV}$ and predicted theoretically by Franta et al. at $\SI{38.79}{meV}$ \cite{Franta2014}. Removal of this feature was accomplished by subtracting the appropriate absorbance spectra, which were taken using ultra-high purity (UHP) undoped Si at each temperature seen in Fig.~\ref{1sT2_1sE_ThermalActivation}. \par 

\begin{figure}[htp]
\includegraphics[width=0.42\textwidth]{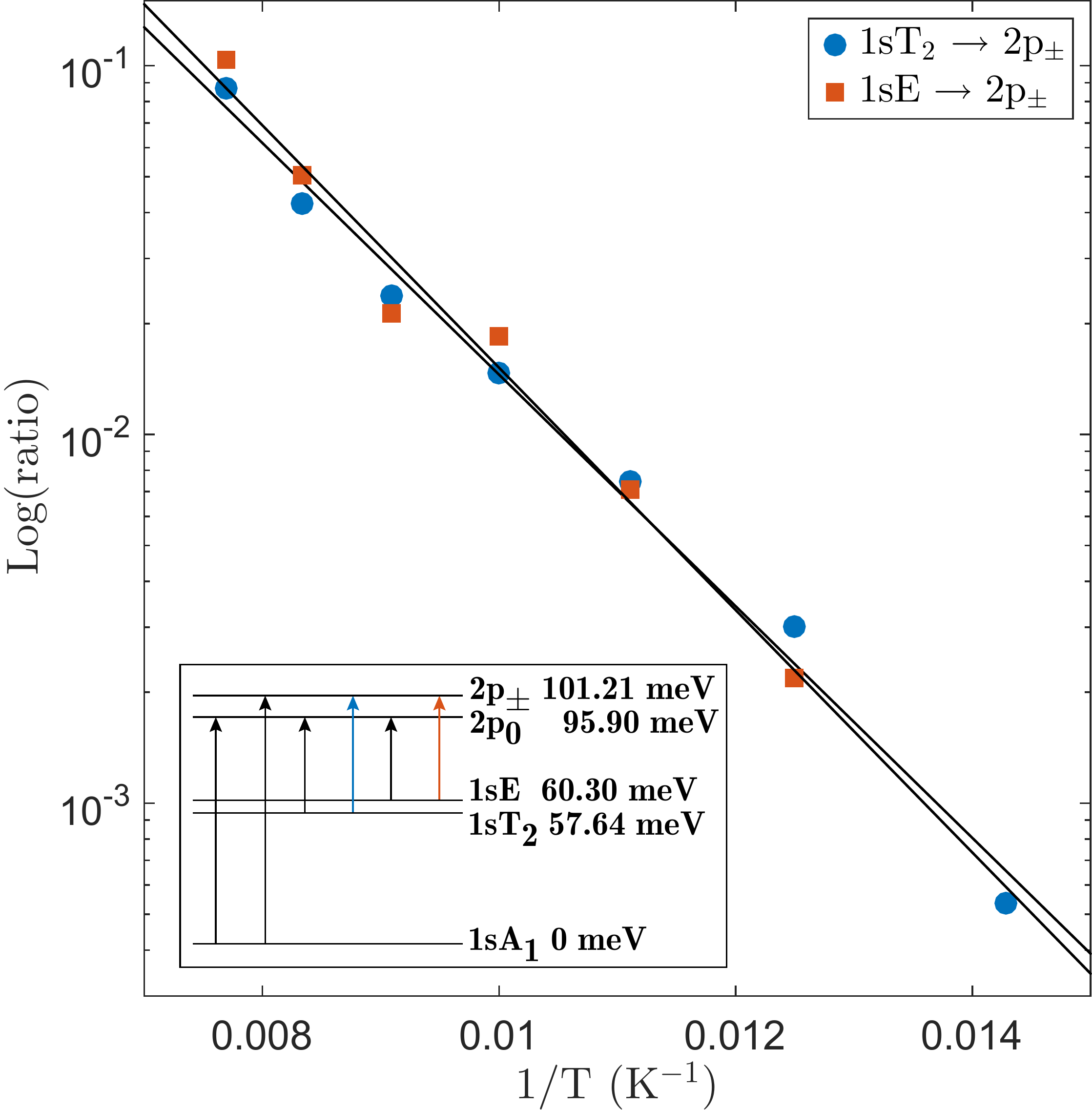}
\caption{Here we show Arrhenius plots corresponding to transitions from valley-orbit states 1sE and 1sT$_2$ to 2p$_{\pm}$. The areas of the 1sE and 1sT$_2$ absorbance features are normalized by the area of the 2p$_0$ absorbance at each temperature. The activation energies extracted from the slopes of these plots are $\SI{62.3}{meV}$ and $\SI{65.2}{meV}$, corresponding to transitions from 1sT$_2$ and 1sE respectively. A level diagram displaying these transitions is included as an inset. Indicated optical spacings relative to the ground state 1sA at $\SI{0}{meV}$ are given as measured at $\SI{100}{K}$.} 
\label{1sT2_1sE_Arrhenius2ppm}
\end{figure}

The integrated intensities of curve fits to the peaks in Fig.~\ref{1sT2_1sE_ThermalActivation} are displayed as a function of inverse temperature in the Arrhenius plots shown in Fig.~\ref{1sT2_1sE_Arrhenius2ppm}, normalized at each temperature by the area of the 2p$_{0}$ transition. These Arrhenius curves correspond to the strongest pair of peaks seen in Fig.~\ref{1sT2_1sE_ThermalActivation}, namely transitions from 1sT$_2$ and 1sE to 2p$_{\pm}$. From the slopes of these we extract activation energies of $\SI{62.3}{meV}$ and $\SI{65.2}{meV}$ for the 1sT$_2$ and 1sE transitions to 2p$_{\pm}$ respectively. We note that these are both slightly higher than the optical spacings, which place 1sT$_2$ and 1sE $\SI{57.64}{meV}$ and $\SI{60.30}{meV}$ respectively above the ground state 1sA. This small disagreement likely results from our inability to measure the sample temperature directly for temperature above $\SI{4.2}{K}$. The observed optical spacings indicate binding energies of $\SI{47.20 \pm 0.01}{meV}$ and $\SI{49.86 \pm 0.01}{meV}$ for 1sE and 1sT$_2$ respectively, in excellent agreement with the preliminary results of Pavlov et al. \cite{Pavlov2018}. \par

We note that there were no signs of similar thermal activation of the valley-orbit excited states for the singly ionized species Mg$_i^+$. This was not unexpected, since the energy differences between 1sA and 1sT$_2$/1sE for the singly ionized species is considerably larger than for Mg$_i^0$. Given the $\SI{256.49}{meV}$ binding energy of the Mg$_i^+$ ground state \cite{Ho1993}, EMT calculations of valley orbit excited state binding energies for double donors by Altarelli \cite{Altarelli1983} suggest spacings of $\SI{101.49}{meV}$($\SI{126.49}{meV}$) between 1sA and 1sT$_2$(1sE) for Mg$_i^+$.

Expanding upon the results of Pavlov et al. \cite{Pavlov2018}, we have demonstrated clear evidence of thermal activation to the valley-orbit levels 1sT$_2$ and 1sE, leading to transitions to both the 2p$_0$ and 2p$_{\pm}$ excited states of Mg$_i^0$. Within a small constant offset, the activation energies extracted from Arrhenius plots associated with the 1sT$_2$ and 1sE to 2p$_{\pm}$ transitions are noted to agree well with the observed optical spacings. No signs of similar transitions from the 1sT$_2$ and 1sE valley orbit excited states of Mg$_i^+$ could be observed.

This work was supported by the Natural Sciences and Engineering Research Council of Canada (NSERC), the Russian Foundation for Basic Research (RFBF Project No. 18-502-12077-DFG), and the Deutsche Forschungsgemeinschaft (DFG No. 389056032). Data available on request from the authors. \par


\begin{thebibliography}{21}%
\makeatletter
\providecommand \@ifxundefined [1]{%
 \@ifx{#1\undefined}
}%
\providecommand \@ifnum [1]{%
 \ifnum #1\expandafter \@firstoftwo
 \else \expandafter \@secondoftwo
 \fi
}%
\providecommand \@ifx [1]{%
 \ifx #1\expandafter \@firstoftwo
 \else \expandafter \@secondoftwo
 \fi
}%
\providecommand \natexlab [1]{#1}%
\providecommand \enquote  [1]{``#1''}%
\providecommand \bibnamefont  [1]{#1}%
\providecommand \bibfnamefont [1]{#1}%
\providecommand \citenamefont [1]{#1}%
\providecommand \href@noop [0]{\@secondoftwo}%
\providecommand \href [0]{\begingroup \@sanitize@url \@href}%
\providecommand \@href[1]{\@@startlink{#1}\@@href}%
\providecommand \@@href[1]{\endgroup#1\@@endlink}%
\providecommand \@sanitize@url [0]{\catcode `\\12\catcode `\$12\catcode
  `\&12\catcode `\#12\catcode `\^12\catcode `\_12\catcode `\%12\relax}%
\providecommand \@@startlink[1]{}%
\providecommand \@@endlink[0]{}%
\providecommand \url  [0]{\begingroup\@sanitize@url \@url }%
\providecommand \@url [1]{\endgroup\@href {#1}{\urlprefix }}%
\providecommand \urlprefix  [0]{URL }%
\providecommand \Eprint [0]{\href }%
\providecommand \doibase [0]{http://dx.doi.org/}%
\providecommand \selectlanguage [0]{\@gobble}%
\providecommand \bibinfo  [0]{\@secondoftwo}%
\providecommand \bibfield  [0]{\@secondoftwo}%
\providecommand \translation [1]{[#1]}%
\providecommand \BibitemOpen [0]{}%
\providecommand \bibitemStop [0]{}%
\providecommand \bibitemNoStop [0]{.\EOS\space}%
\providecommand \EOS [0]{\spacefactor3000\relax}%
\providecommand \BibitemShut  [1]{\csname bibitem#1\endcsname}%
\let\auto@bib@innerbib\@empty
\bibitem [{\citenamefont {Franks}\ and\ \citenamefont
  {Robertson}(1967)}]{Franks1967}%
  \BibitemOpen
  \bibfield  {author} {\bibinfo {author} {\bibfnamefont {R.}~\bibnamefont
  {Franks}}\ and\ \bibinfo {author} {\bibfnamefont {J.}~\bibnamefont
  {Robertson}},\ }\href {\doibase 10.1016/0038-1098(67)90598-4} {\bibfield
  {journal} {\bibinfo  {journal} {Solid State Communications}\ }\textbf
  {\bibinfo {volume} {5}},\ \bibinfo {pages} {479 } (\bibinfo {year}
  {1967})}\BibitemShut {NoStop}%
\bibitem [{\citenamefont {Ho}\ and\ \citenamefont {Ramdas}(1972)}]{Ho1972}%
  \BibitemOpen
  \bibfield  {author} {\bibinfo {author} {\bibfnamefont {L.~T.}\ \bibnamefont
  {Ho}}\ and\ \bibinfo {author} {\bibfnamefont {A.~K.}\ \bibnamefont
  {Ramdas}},\ }\href {\doibase 10.1103/PhysRevB.5.462} {\bibfield  {journal}
  {\bibinfo  {journal} {Phys. Rev. B}\ }\textbf {\bibinfo {volume} {5}},\
  \bibinfo {pages} {462} (\bibinfo {year} {1972})}\BibitemShut {NoStop}%
\bibitem [{\citenamefont {Ho}\ \emph {et~al.}(1993)\citenamefont {Ho},
  \citenamefont {Lin},\ and\ \citenamefont {Lin}}]{Ho1993}%
  \BibitemOpen
  \bibfield  {author} {\bibinfo {author} {\bibfnamefont {L.~T.}\ \bibnamefont
  {Ho}}, \bibinfo {author} {\bibfnamefont {F.~Y.}\ \bibnamefont {Lin}}, \ and\
  \bibinfo {author} {\bibfnamefont {W.~J.}\ \bibnamefont {Lin}},\ }\href
  {\doibase 10.1007/BF02084585} {\bibfield  {journal} {\bibinfo  {journal}
  {International Journal of Infrared and Millimeter Waves}\ }\textbf {\bibinfo
  {volume} {14}},\ \bibinfo {pages} {1099} (\bibinfo {year}
  {1993})}\BibitemShut {NoStop}%
\bibitem [{\citenamefont {Ho}(1998)}]{Ho1998}%
  \BibitemOpen
  \bibfield  {author} {\bibinfo {author} {\bibfnamefont {L.~T.}\ \bibnamefont
  {Ho}},\ }\href {\doibase
  10.1002/(SICI)1521-3951(199812)210:2<313::AID-PSSB313>3.0.CO;2-C} {\bibfield
  {journal} {\bibinfo  {journal} {physica status solidi (b)}\ }\textbf
  {\bibinfo {volume} {210}},\ \bibinfo {pages} {313} (\bibinfo {year}
  {1998})}\BibitemShut {NoStop}%
\bibitem [{\citenamefont {Ho}(2003{\natexlab{a}})}]{Ho2003}%
  \BibitemOpen
  \bibfield  {author} {\bibinfo {author} {\bibfnamefont {L.~T.}\ \bibnamefont
  {Ho}},\ }\href {\doibase 10.1002/pssc.200306203} {\bibfield  {journal}
  {\bibinfo  {journal} {physica status solidi (c)}\ }\textbf {\bibinfo {volume}
  {0}},\ \bibinfo {pages} {721} (\bibinfo {year}
  {2003}{\natexlab{a}})}\BibitemShut {NoStop}%
\bibitem [{\citenamefont {Ho}(2003{\natexlab{b}})}]{2Ho2003}%
  \BibitemOpen
  \bibfield  {author} {\bibinfo {author} {\bibfnamefont {L.~T.}\ \bibnamefont
  {Ho}},\ }in\ \href {\doibase 10.4028/www.scientific.net/DDF.221-223.41}
  {\emph {\bibinfo {booktitle} {Defects and Diffusion in Semiconductors,
  2003}}},\ \bibinfo {series} {Defect and Diffusion Forum}, Vol.\ \bibinfo
  {volume} {221}\ (\bibinfo  {publisher} {Trans Tech Publications},\ \bibinfo
  {year} {2003})\ pp.\ \bibinfo {pages} {41--50}\BibitemShut {NoStop}%
\bibitem [{\citenamefont {Ho}(2006)}]{Ho2006}%
  \BibitemOpen
  \bibfield  {author} {\bibinfo {author} {\bibfnamefont {L.~T.}\ \bibnamefont
  {Ho}},\ }\href {\doibase 10.1016/j.physb.2005.12.041} {\bibfield  {journal}
  {\bibinfo  {journal} {Physica B: Condensed Matter}\ }\textbf {\bibinfo
  {volume} {376-377}},\ \bibinfo {pages} {154 } (\bibinfo {year}
  {2006})}\BibitemShut {NoStop}%
\bibitem [{\citenamefont {Abraham}\ \emph
  {et~al.}(2018{\natexlab{a}})\citenamefont {Abraham}, \citenamefont {DeAbreu},
  \citenamefont {Morse}, \citenamefont {Shuman}, \citenamefont {Portsel},
  \citenamefont {Lodygin}, \citenamefont {Astrov}, \citenamefont {Abrosimov},
  \citenamefont {Pavlov}, \citenamefont {H\"ubers}, \citenamefont {Simmons},\
  and\ \citenamefont {Thewalt}}]{Abraham2018}%
  \BibitemOpen
  \bibfield  {author} {\bibinfo {author} {\bibfnamefont {R.~J.~S.}\
  \bibnamefont {Abraham}}, \bibinfo {author} {\bibfnamefont {A.}~\bibnamefont
  {DeAbreu}}, \bibinfo {author} {\bibfnamefont {K.~J.}\ \bibnamefont {Morse}},
  \bibinfo {author} {\bibfnamefont {V.~B.}\ \bibnamefont {Shuman}}, \bibinfo
  {author} {\bibfnamefont {L.~M.}\ \bibnamefont {Portsel}}, \bibinfo {author}
  {\bibfnamefont {A.~N.}\ \bibnamefont {Lodygin}}, \bibinfo {author}
  {\bibfnamefont {{\relax Yu}.~A.}\ \bibnamefont {Astrov}}, \bibinfo {author}
  {\bibfnamefont {N.~V.}\ \bibnamefont {Abrosimov}}, \bibinfo {author}
  {\bibfnamefont {S.~G.}\ \bibnamefont {Pavlov}}, \bibinfo {author}
  {\bibfnamefont {H.-W.}\ \bibnamefont {H\"ubers}}, \bibinfo {author}
  {\bibfnamefont {S.}~\bibnamefont {Simmons}}, \ and\ \bibinfo {author}
  {\bibfnamefont {M.~L.~W.}\ \bibnamefont {Thewalt}},\ }\href {\doibase
  10.1103/PhysRevB.98.045202} {\bibfield  {journal} {\bibinfo  {journal} {Phys.
  Rev. B}\ }\textbf {\bibinfo {volume} {98}},\ \bibinfo {pages} {045202}
  (\bibinfo {year} {2018}{\natexlab{a}})}\BibitemShut {NoStop}%
\bibitem [{\citenamefont {Abraham}\ \emph
  {et~al.}(2018{\natexlab{b}})\citenamefont {Abraham}, \citenamefont {DeAbreu},
  \citenamefont {Morse}, \citenamefont {Shuman}, \citenamefont {Portsel},
  \citenamefont {Lodygin}, \citenamefont {Astrov}, \citenamefont {Abrosimov},
  \citenamefont {Pavlov}, \citenamefont {H\"ubers}, \citenamefont {Simmons},\
  and\ \citenamefont {Thewalt}}]{2Abraham2018}%
  \BibitemOpen
  \bibfield  {author} {\bibinfo {author} {\bibfnamefont {R.~J.~S.}\
  \bibnamefont {Abraham}}, \bibinfo {author} {\bibfnamefont {A.}~\bibnamefont
  {DeAbreu}}, \bibinfo {author} {\bibfnamefont {K.~J.}\ \bibnamefont {Morse}},
  \bibinfo {author} {\bibfnamefont {V.~B.}\ \bibnamefont {Shuman}}, \bibinfo
  {author} {\bibfnamefont {L.~M.}\ \bibnamefont {Portsel}}, \bibinfo {author}
  {\bibfnamefont {A.~N.}\ \bibnamefont {Lodygin}}, \bibinfo {author}
  {\bibfnamefont {{\relax Yu}.~A.}\ \bibnamefont {Astrov}}, \bibinfo {author}
  {\bibfnamefont {N.~V.}\ \bibnamefont {Abrosimov}}, \bibinfo {author}
  {\bibfnamefont {S.~G.}\ \bibnamefont {Pavlov}}, \bibinfo {author}
  {\bibfnamefont {H.-W.}\ \bibnamefont {H\"ubers}}, \bibinfo {author}
  {\bibfnamefont {S.}~\bibnamefont {Simmons}}, \ and\ \bibinfo {author}
  {\bibfnamefont {M.~L.~W.}\ \bibnamefont {Thewalt}},\ }\href {\doibase
  10.1103/PhysRevB.98.205203} {\bibfield  {journal} {\bibinfo  {journal} {Phys.
  Rev. B}\ }\textbf {\bibinfo {volume} {98}},\ \bibinfo {pages} {205203}
  (\bibinfo {year} {2018}{\natexlab{b}})}\BibitemShut {NoStop}%
\bibitem [{\citenamefont {G.~Pavlov}\ \emph {et~al.}(2018)\citenamefont
  {G.~Pavlov}, \citenamefont {V.~Abrosimov}, \citenamefont {B.~Shuman},
  \citenamefont {Portsel}, \citenamefont {N.~Lodygin}, \citenamefont
  {A.~Astrov}, \citenamefont {{\relax Kh}.~Zhukavin}, \citenamefont
  {N.~Shastin}, \citenamefont {Irmscher}, \citenamefont {Pohl},\ and\
  \citenamefont {H\"ubers}}]{Pavlov2018}%
  \BibitemOpen
  \bibfield  {author} {\bibinfo {author} {\bibfnamefont {S.}~\bibnamefont
  {G.~Pavlov}}, \bibinfo {author} {\bibfnamefont {N.}~\bibnamefont
  {V.~Abrosimov}}, \bibinfo {author} {\bibfnamefont {V.}~\bibnamefont
  {B.~Shuman}}, \bibinfo {author} {\bibfnamefont {L.~M.}\ \bibnamefont
  {Portsel}}, \bibinfo {author} {\bibfnamefont {{\relax A}.}~\bibnamefont
  {N.~Lodygin}}, \bibinfo {author} {\bibfnamefont {{\relax Yu}.}~\bibnamefont
  {A.~Astrov}}, \bibinfo {author} {\bibfnamefont {R.}~\bibnamefont {{\relax
  Kh}.~Zhukavin}}, \bibinfo {author} {\bibfnamefont {V.}~\bibnamefont
  {N.~Shastin}}, \bibinfo {author} {\bibfnamefont {K.}~\bibnamefont
  {Irmscher}}, \bibinfo {author} {\bibfnamefont {A.}~\bibnamefont {Pohl}}, \
  and\ \bibinfo {author} {\bibfnamefont {H.-W.}\ \bibnamefont {H\"ubers}},\
  }\href {\doibase 10.1002/pssb.201800514} {\bibfield  {journal} {\bibinfo
  {journal} {physica status solidi (b)}\ } (\bibinfo {year} {2018}),\
  10.1002/pssb.201800514}\BibitemShut {NoStop}%
\bibitem [{\citenamefont {H\"ubers}\ \emph {et~al.}(2005)\citenamefont
  {H\"ubers}, \citenamefont {Pavlov},\ and\ \citenamefont
  {Shastin}}]{Hubers2005}%
  \BibitemOpen
  \bibfield  {author} {\bibinfo {author} {\bibfnamefont {H.-W.}\ \bibnamefont
  {H\"ubers}}, \bibinfo {author} {\bibfnamefont {S.~G.}\ \bibnamefont
  {Pavlov}}, \ and\ \bibinfo {author} {\bibfnamefont {V.~N.}\ \bibnamefont
  {Shastin}},\ }\href {\doibase 10.1088/0268-1242/20/7/011} {\bibfield
  {journal} {\bibinfo  {journal} {Semiconductor Science and Technology}\
  }\textbf {\bibinfo {volume} {20}},\ \bibinfo {pages} {S211} (\bibinfo {year}
  {2005})}\BibitemShut {NoStop}%
\bibitem [{\citenamefont {{Shastin}}\ \emph {et~al.}(2019)\citenamefont
  {{Shastin}}, \citenamefont {{Zhukavin}}, \citenamefont {{Kovalevsky}},
  \citenamefont {{Tsyplenkov}}, \citenamefont {{Rumyantsev}}, \citenamefont
  {{Shengurov}}, \citenamefont {{Pavlov}}, \citenamefont {{Shuman}},
  \citenamefont {{Portsel}}, \citenamefont {{Lodygin}}, \citenamefont
  {{Astrov}}, \citenamefont {{Abrosimov}}, \citenamefont {{Klopf}},\ and\
  \citenamefont {{H{\"u}bers}}}]{Shastin2019}%
  \BibitemOpen
  \bibfield  {author} {\bibinfo {author} {\bibfnamefont {V.~N.}\ \bibnamefont
  {{Shastin}}}, \bibinfo {author} {\bibfnamefont {R.~{\relax Kh}.}\
  \bibnamefont {{Zhukavin}}}, \bibinfo {author} {\bibfnamefont {K.~A.}\
  \bibnamefont {{Kovalevsky}}}, \bibinfo {author} {\bibfnamefont {V.~V.}\
  \bibnamefont {{Tsyplenkov}}}, \bibinfo {author} {\bibfnamefont {V.~V.}\
  \bibnamefont {{Rumyantsev}}}, \bibinfo {author} {\bibfnamefont {D.~V.}\
  \bibnamefont {{Shengurov}}}, \bibinfo {author} {\bibfnamefont {S.~G.}\
  \bibnamefont {{Pavlov}}}, \bibinfo {author} {\bibfnamefont {V.~B.}\
  \bibnamefont {{Shuman}}}, \bibinfo {author} {\bibfnamefont {L.~M.}\
  \bibnamefont {{Portsel}}}, \bibinfo {author} {\bibfnamefont {A.~N.}\
  \bibnamefont {{Lodygin}}}, \bibinfo {author} {\bibfnamefont {{\relax
  Yu}.~A.}\ \bibnamefont {{Astrov}}}, \bibinfo {author} {\bibfnamefont {N.~V.}\
  \bibnamefont {{Abrosimov}}}, \bibinfo {author} {\bibfnamefont {J.~M.}\
  \bibnamefont {{Klopf}}}, \ and\ \bibinfo {author} {\bibfnamefont {H.-W.}\
  \bibnamefont {{H{\"u}bers}}},\ }\href {\doibase 10.1134/S1063782619090197}
  {\bibfield  {journal} {\bibinfo  {journal} {Semiconductors}\ }\textbf
  {\bibinfo {volume} {53}},\ \bibinfo {pages} {1234} (\bibinfo {year}
  {2019})}\BibitemShut {NoStop}%
\bibitem [{\citenamefont {Ascarelli}\ and\ \citenamefont
  {Rodriguez}(1961)}]{Ascarelli1961}%
  \BibitemOpen
  \bibfield  {author} {\bibinfo {author} {\bibfnamefont {G.}~\bibnamefont
  {Ascarelli}}\ and\ \bibinfo {author} {\bibfnamefont {S.}~\bibnamefont
  {Rodriguez}},\ }\href {\doibase 10.1103/PhysRev.124.1321} {\bibfield
  {journal} {\bibinfo  {journal} {Phys. Rev.}\ }\textbf {\bibinfo {volume}
  {124}},\ \bibinfo {pages} {1321} (\bibinfo {year} {1961})}\BibitemShut
  {NoStop}%
\bibitem [{\citenamefont {Lax}(1960)}]{Lax1960}%
  \BibitemOpen
  \bibfield  {author} {\bibinfo {author} {\bibfnamefont {M.}~\bibnamefont
  {Lax}},\ }\href {\doibase 10.1103/PhysRev.119.1502} {\bibfield  {journal}
  {\bibinfo  {journal} {Phys. Rev.}\ }\textbf {\bibinfo {volume} {119}},\
  \bibinfo {pages} {1502} (\bibinfo {year} {1960})}\BibitemShut {NoStop}%
\bibitem [{\citenamefont {Morse}\ \emph {et~al.}(2017)\citenamefont {Morse},
  \citenamefont {Abraham}, \citenamefont {DeAbreu}, \citenamefont {Bowness},
  \citenamefont {Richards}, \citenamefont {Riemann}, \citenamefont {Abrosimov},
  \citenamefont {Becker}, \citenamefont {Pohl}, \citenamefont {Thewalt},\ and\
  \citenamefont {Simmons}}]{Morse2017}%
  \BibitemOpen
  \bibfield  {author} {\bibinfo {author} {\bibfnamefont {K.~J.}\ \bibnamefont
  {Morse}}, \bibinfo {author} {\bibfnamefont {R.~J.~S.}\ \bibnamefont
  {Abraham}}, \bibinfo {author} {\bibfnamefont {A.}~\bibnamefont {DeAbreu}},
  \bibinfo {author} {\bibfnamefont {C.}~\bibnamefont {Bowness}}, \bibinfo
  {author} {\bibfnamefont {T.~S.}\ \bibnamefont {Richards}}, \bibinfo {author}
  {\bibfnamefont {H.}~\bibnamefont {Riemann}}, \bibinfo {author} {\bibfnamefont
  {N.~V.}\ \bibnamefont {Abrosimov}}, \bibinfo {author} {\bibfnamefont
  {P.}~\bibnamefont {Becker}}, \bibinfo {author} {\bibfnamefont {H.-J.}\
  \bibnamefont {Pohl}}, \bibinfo {author} {\bibfnamefont {M.~L.~W.}\
  \bibnamefont {Thewalt}}, \ and\ \bibinfo {author} {\bibfnamefont
  {S.}~\bibnamefont {Simmons}},\ }\href {\doibase 10.1126/sciadv.1700930}
  {\bibfield  {journal} {\bibinfo  {journal} {Science Advances}\ }\textbf
  {\bibinfo {volume} {3}},\ \bibinfo {pages} {e1700930} (\bibinfo {year}
  {2017})}\BibitemShut {NoStop}%
\bibitem [{\citenamefont {Steger}\ \emph {et~al.}(2009)\citenamefont {Steger},
  \citenamefont {Yang}, \citenamefont {Thewalt}, \citenamefont {Cardona},
  \citenamefont {Riemann}, \citenamefont {Abrosimov}, \citenamefont
  {Churbanov}, \citenamefont {Gusev}, \citenamefont {Bulanov}, \citenamefont
  {Kovalev}, \citenamefont {Kaliteevskii}, \citenamefont {Godisov},
  \citenamefont {Becker}, \citenamefont {Pohl}, \citenamefont {Haller},\ and\
  \citenamefont {Ager}}]{Steger2009}%
  \BibitemOpen
  \bibfield  {author} {\bibinfo {author} {\bibfnamefont {M.}~\bibnamefont
  {Steger}}, \bibinfo {author} {\bibfnamefont {A.}~\bibnamefont {Yang}},
  \bibinfo {author} {\bibfnamefont {M.~L.~W.}\ \bibnamefont {Thewalt}},
  \bibinfo {author} {\bibfnamefont {M.}~\bibnamefont {Cardona}}, \bibinfo
  {author} {\bibfnamefont {H.}~\bibnamefont {Riemann}}, \bibinfo {author}
  {\bibfnamefont {N.~V.}\ \bibnamefont {Abrosimov}}, \bibinfo {author}
  {\bibfnamefont {M.~F.}\ \bibnamefont {Churbanov}}, \bibinfo {author}
  {\bibfnamefont {A.~V.}\ \bibnamefont {Gusev}}, \bibinfo {author}
  {\bibfnamefont {A.~D.}\ \bibnamefont {Bulanov}}, \bibinfo {author}
  {\bibfnamefont {I.~D.}\ \bibnamefont {Kovalev}}, \bibinfo {author}
  {\bibfnamefont {A.~K.}\ \bibnamefont {Kaliteevskii}}, \bibinfo {author}
  {\bibfnamefont {O.~N.}\ \bibnamefont {Godisov}}, \bibinfo {author}
  {\bibfnamefont {P.}~\bibnamefont {Becker}}, \bibinfo {author} {\bibfnamefont
  {H.-J.}\ \bibnamefont {Pohl}}, \bibinfo {author} {\bibfnamefont {E.~E.}\
  \bibnamefont {Haller}}, \ and\ \bibinfo {author} {\bibfnamefont {J.~W.}\
  \bibnamefont {Ager}},\ }\href {\doibase 10.1103/PhysRevB.80.115204}
  {\bibfield  {journal} {\bibinfo  {journal} {Phys. Rev. B}\ }\textbf {\bibinfo
  {volume} {80}},\ \bibinfo {pages} {115204} (\bibinfo {year}
  {2009})}\BibitemShut {NoStop}%
\bibitem [{\citenamefont {Mayur}\ \emph {et~al.}(1993)\citenamefont {Mayur},
  \citenamefont {Sciacca}, \citenamefont {Ramdas},\ and\ \citenamefont
  {Rodriguez}}]{Mayur1993}%
  \BibitemOpen
  \bibfield  {author} {\bibinfo {author} {\bibfnamefont {A.~J.}\ \bibnamefont
  {Mayur}}, \bibinfo {author} {\bibfnamefont {M.~D.}\ \bibnamefont {Sciacca}},
  \bibinfo {author} {\bibfnamefont {A.~K.}\ \bibnamefont {Ramdas}}, \ and\
  \bibinfo {author} {\bibfnamefont {S.}~\bibnamefont {Rodriguez}},\ }\href
  {\doibase 10.1103/PhysRevB.48.10893} {\bibfield  {journal} {\bibinfo
  {journal} {Phys. Rev. B}\ }\textbf {\bibinfo {volume} {48}},\ \bibinfo
  {pages} {10893} (\bibinfo {year} {1993})}\BibitemShut {NoStop}%
\bibitem [{\citenamefont {Shuman}\ \emph
  {et~al.}(2017{\natexlab{a}})\citenamefont {Shuman}, \citenamefont {Astrov},
  \citenamefont {Lodygin},\ and\ \citenamefont {Portsel}}]{1Shuman2017}%
  \BibitemOpen
  \bibfield  {author} {\bibinfo {author} {\bibfnamefont {V.~B.}\ \bibnamefont
  {Shuman}}, \bibinfo {author} {\bibfnamefont {{\relax Yu}.~A.}\ \bibnamefont
  {Astrov}}, \bibinfo {author} {\bibfnamefont {A.~N.}\ \bibnamefont {Lodygin}},
  \ and\ \bibinfo {author} {\bibfnamefont {L.~M.}\ \bibnamefont {Portsel}},\
  }\href {\doibase 10.1134/S1063782617080292} {\bibfield  {journal} {\bibinfo
  {journal} {Semiconductors}\ }\textbf {\bibinfo {volume} {51}},\ \bibinfo
  {pages} {1031} (\bibinfo {year} {2017}{\natexlab{a}})}\BibitemShut {NoStop}%
\bibitem [{\citenamefont {Shuman}\ \emph
  {et~al.}(2017{\natexlab{b}})\citenamefont {Shuman}, \citenamefont
  {Lavrent'ev}, \citenamefont {Astrov}, \citenamefont {Lodygin},\ and\
  \citenamefont {Portsel}}]{2Shuman2017}%
  \BibitemOpen
  \bibfield  {author} {\bibinfo {author} {\bibfnamefont {V.~B.}\ \bibnamefont
  {Shuman}}, \bibinfo {author} {\bibfnamefont {A.~A.}\ \bibnamefont
  {Lavrent'ev}}, \bibinfo {author} {\bibfnamefont {{\relax Yu}.~A.}\
  \bibnamefont {Astrov}}, \bibinfo {author} {\bibfnamefont {A.~N.}\
  \bibnamefont {Lodygin}}, \ and\ \bibinfo {author} {\bibfnamefont {L.~M.}\
  \bibnamefont {Portsel}},\ }\href {\doibase 10.1134/S1063782617010237}
  {\bibfield  {journal} {\bibinfo  {journal} {Semiconductors}\ }\textbf
  {\bibinfo {volume} {51}},\ \bibinfo {pages} {1} (\bibinfo {year}
  {2017}{\natexlab{b}})}\BibitemShut {NoStop}%
\bibitem [{\citenamefont {Franta}\ \emph {et~al.}(2014)\citenamefont {Franta},
  \citenamefont {Ne\v{c}as}, \citenamefont {Zaj\'{i}\v{c}kov\'{a}},\ and\
  \citenamefont {Ohl\'{i}dal}}]{Franta2014}%
  \BibitemOpen
  \bibfield  {author} {\bibinfo {author} {\bibfnamefont {D.}~\bibnamefont
  {Franta}}, \bibinfo {author} {\bibfnamefont {D.}~\bibnamefont {Ne\v{c}as}},
  \bibinfo {author} {\bibfnamefont {L.}~\bibnamefont {Zaj\'{i}\v{c}kov\'{a}}},
  \ and\ \bibinfo {author} {\bibfnamefont {I.}~\bibnamefont {Ohl\'{i}dal}},\
  }\href {\doibase 10.1364/OME.4.001641} {\bibfield  {journal} {\bibinfo
  {journal} {Opt. Mater. Express}\ }\textbf {\bibinfo {volume} {4}},\ \bibinfo
  {pages} {1641} (\bibinfo {year} {2014})}\BibitemShut {NoStop}%
\bibitem [{\citenamefont {{Altarelli}}(1983)}]{Altarelli1983}%
  \BibitemOpen
  \bibfield  {author} {\bibinfo {author} {\bibfnamefont {M.}~\bibnamefont
  {{Altarelli}}},\ }\href {\doibase 10.1016/0378-4363(83)90459-X} {\bibfield
  {journal} {\bibinfo  {journal} {Physica B+C}\ }\textbf {\bibinfo {volume}
  {117}},\ \bibinfo {pages} {122} (\bibinfo {year} {1983})}\BibitemShut
  {NoStop}%
\end{thebibliography}
\end{document}